\documentclass[twocolumn,reprint,amsmath,amssymb,prb]{revtex4-1}

\usepackage{docs}
\usepackage{epsfig}
\usepackage{graphicx}
\usepackage{bm}%
\usepackage[colorlinks=true,linkcolor=blue]{hyperref}%

\begin{document}

\title{Complex electronic structure of Ca$_{1-x}$Sr$_x$RuO$_3$}

\author{Kalobaran Maiti}
\altaffiliation{Corresponding author: kbmaiti@tifr.res.in}

\author{Ravi Shankar Singh}
\altaffiliation{Present Address: Indian Institute of Science
Education and Research, Bhopal ITI (Gas Rahat) Building, Govindpura,
Bhopal - 462 023, India}

\affiliation{Department of Condensed Matter Physics and Materials
Science, Tata Institute of Fundamental Research, Homi Bhabha Road,
Colaba, Mumbai - 400 005, INDIA}

\date{\today}

\begin{abstract}

We investigate the core level spectra of Ca$_{1-x}$Sr$_x$RuO$_3$
employing high resolution photoemission spectroscopy. Sample surface
appears to be dominated by the contributions from Ru-O layers. Sr
3$p$ core level spectra are sharp and asymmetric in SrRuO$_3$ as
expected in a metallic system, and exhibit multiple features for the
intermediate compositions that can be attributed to the difference
in Ca-O and Sr-O covalency. The Ru core level spectra exhibit
distinct signature of satellite features due to the finite electron
correlations strength among Ru 4$d$ electrons. The intensity of the
satellite feature is weaker in the surface spectra compared to the
bulk. The low temperature spectra exhibit enhancement of satellite
intensity in the spectra corresponding to ferromagnetic compositions
due to the inter-site exchange coupling induced depletion of the
intensity at the Fermi level. The increase in $x$ leads to a
decrease in satellite intensity that has been attributed to the
increase in hopping interaction strength due to the enhancement of
the Ru-O-Ru bond angle. Evidently, the complex electronic properties
of these materials are derived from the interplay between the
electron correlation and hopping interaction strengths.

\end{abstract}


\maketitle

\section{Introduction}

4$d$ and 5$d$ transition metal oxides have drawn significant
attention during last few decades due to the discovery of many
interesting properties arising from the competition between
electron-electron Coulomb repulsion (termed as `\emph{electron
correlation}') induced effects, spin-orbit coupling and large
crystal field effects. While 5$d$ oxides exhibit signature of
unusual transport and magnetic
properties,\cite{bairo3,sr2iro4,y2ir2o7} 4$d$ transition metal
oxides, in particular Ru-based oxides exhibit plethora of
interesting properties such as superconductivity,\cite{sr2ruo4}
non-Fermi liquid behavior,\cite{nfl,klein} unusual magnetic ground
states\cite{nfl,klein,rss,cao} {\it etc.} Even a simple perovskite
compound, SrRuO$_3$ exhibits ferromagnetic long range order (Curie
temperature, $T_C$ $\sim$ 160~K) despite having highly extended 4$d$
character of the valence electrons.\cite{rss,cao} Another
isostructural compound, CaRuO$_3$ exhibit complex ground state
properties,\cite{rana} with controversies on the existence of
antiferromagnetic order,\cite{vidya,callaghan,longo,sugiyama} or the
absence of any long-range order down to the lowest temperature
studied.\cite{nfl,klein,cao,martinez} The behavior of CaRuO$_3$ is
often discussed considering proximity to the quantum
criticality.\cite{nfl,klein}

Extensive studies have been carried out on these materials in the
form of bulk samples and thin films. Photoemission studies suggest
significantly different surface and bulk electronic
structure,\cite{prbr-ravi,takizawa-prbr-2005} signature of
particle-hole asymmetry,\cite{epl-ravi} unusual thermal evolution of
local structural disorder,\cite{prb-debdutta} unusual multiple
features in the core level spectra of alkaline earth materials and
oxygen as well,\cite{spec-core} etc. In these systems, the
conduction electrons consists of hybridized Ru 4$d$ and O 2$p$
states and moves via corner shared RuO$_6$ network. Hence, the
tilting and buckling of the RuO$_6$
octahedra\cite{ramarao,nakatsugawa,kobayashi} plays dominant role in
deriving the electronic properties of this system. The Ru-O-Ru bond
angles in SrRuO$_3$ are 167.6$^\circ$ and 159.7$^\circ$, while those
in CaRuO$_3$ are 149.6$^\circ$ - 149.8$^\circ$. The electron hopping
interaction strength and therefore the valence band width, $W$ is
expected to be larger in SrRuO$_3$ than that in CaRuO$_3$. Thus, the
effective electron correlation strength, $U/W$ ($U$ =
electron-electron Coulomb repulsion strength) is expected to be
higher in CaRuO$_3$ than that in SrRuO$_3$. This difference in $U/W$
is often considered as the origin of the difference in ground state
properties between these compounds, while some studies predicted
similar $U/W$ in these compounds. Evidently, the electronic
properties in these compounds still remain to be a puzzle despite
numerous studies. Here, we studied core level spectra of the whole
series of compounds with formula, Ca$_{1-x}$Sr$_x$RuO$_3$ employing
high resolution $x$-ray photoemission spectroscopy (XPS). Our
results suggest interesting spectral evolution with temperature and
composition, which could provide a clue to the understanding of the
complex behavior of these materials.

\section{Experimental}

The samples of Ca$_{1-x}$Sr$_{x}$RuO$_{3}$ (for \emph{x}=0.0, 0.2,
0.5, 0.7, 1.0) were prepared by conventional solid state reaction
rout. High purity ingredients, CaCO$_{3}$, SrCO$_{3}$ and RuO$_{2}$
were mixed in appropriate molar concentration, ground together for
about an hour and calcined at 1273 K for about 24 hours and then
treated at 1523 K. To achieve large grain size, the samples were
sintered for about 72 hours in pellet form at the preparation
temperature of 1523~K with two intermittent grounding. $X$-ray
diffraction (XRD) patterns were recorded using Cu $K\alpha$ line at
every step. XRD patterns show orthorhombic crystal structure for all
the samples with lattice constants similar to the observed values
for single crystalline samples. It is to note here that the size of
the single crystals are too small to carry out photoemission
spectroscopy. It was almost impossible to grow samples of suitable
size for the photoemission measurements and hence, highly sintered
polycrystalline samples were used for these studies.

Direct current magnetic susceptibility measurements were performed
using vibrating sample magnetometer (VSM). SrRuO$_3$ shows
ferromagnetic transition at 165~K. Magnetic susceptibility, of
Ca$_{1-x}$Sr$_{x}$RuO$_{3}$ reveals a broad ferromagnetic transition
close to 150 K for all values of $x \geq$ 0.2 [Ref. \cite{rss}].
$\mu$$_{eff}$ for all the samples in the paramagnetic region has
been found to be close to the theoretical spin only value of
2.83$\mu$$_{B}$ corresponding to $t_{2g\uparrow}^3
t_{2g\downarrow}^1$ occupation in Ru$^{4+}$.

Photoemission measurements were performed on \emph{in situ} (base
pressure better than 4$\times10^{-11}$ torr) scraped samples using
an SES2002 Gammadata Scienta analyzer. Reproducibility of the
photoemission spectra and cleanliness of the sample surface was
confirmed at each scrapping trial. A high purity silver was also
mounted on the same sample holder in electrical contact with other
samples to determine the Fermi level. Measurements were carried out
using different photon sources, at different temperatures. Total
instrumental resolution was fixed to 0.8~eV for Mg $K\alpha$ and
0.3~eV for monochromatic Al $K\alpha$ photon energies.

\section{Results and discussions}

\begin{figure}
 \centerline{\epsfysize=5.0in \epsffile{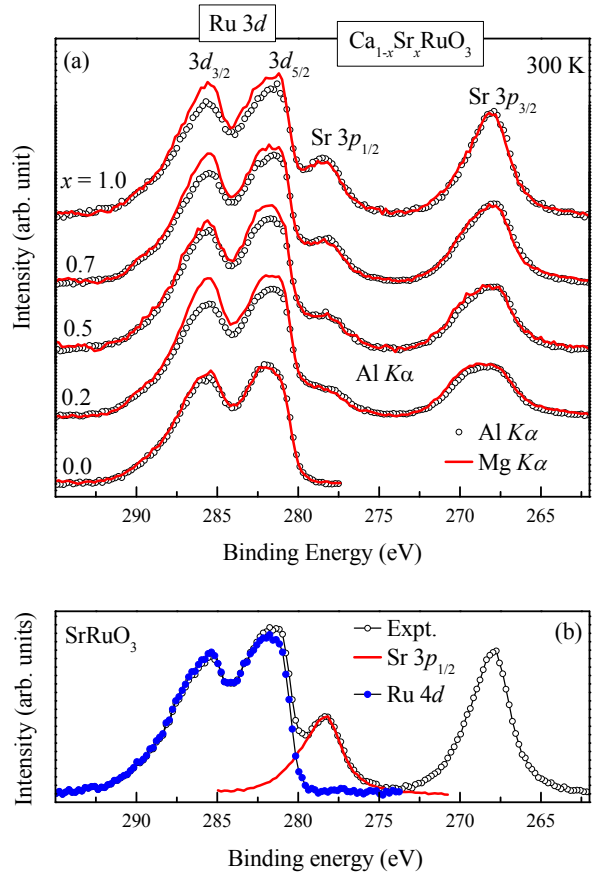}}
\vspace{-2ex}
 \caption{(a) Ru 3$d$ and Sr 3$p$ core level spectra in Ca$_{1-x}$Sr$_x$RuO$_3$ for
different values of $x$ using Al $K\alpha$ (open circles) and Mg
$k\alpha$ (solid lines) radiations at 300 K. (b) Ru 3$d$ spectrum
(closed circles) is extracted from the experimental spectrum (open
circles) of SrRuO$_3$ after subtracting the Sr 3$p_{1/2}$
contributions (solid line).}
 \vspace{-2ex}
\end{figure}

In Fig. 1, we show the spectral region for Ru 3$d$ and Sr 3$p$ core
levels of Ca$_{1-x}$Sr$_x$RuO$_3$ for different values of $x$
obtained using Al $K\alpha$ (circles) and Mg $K\alpha$ (solid lines)
radiations at room temperature. Each spectrum exhibit multiple
features - Sr 3$p$ spin-orbit split signals appear around 268 eV and
278.5 eV binding energies indicating a spin-orbit splitting of about
10.5 eV for Sr 3$p$ levels. Ru 3$d$ features appear around 282 eV
and 285.5 eV binding energies with a spin-orbit splitting of about
3.5 eV for Ru 3$d$ levels. In order to investigate the Sr 3$p_{1/2}$
contributions at the lower binding energy tail of the Ru 3$d$ level,
we have simulated the Sr 3$p_{1/2}$ signal taking the experimental
Sr 3$p_{3/2}$ signal and the multiplicity ratio of 1:2. Here, the
lineshape of the spectral function is kept same, which is a good
approximation for such deep core levels and found to work well in
the present case. The extraction procedure is demonstrated in Fig.
1(b) for a typical spectrum of SrRuO$_3$ exhibiting Ru 3$d$ signal
similar to that of CaRuO$_3$, where there is no such signal present.

The intensity of Sr 3$p$ signal with respect to the Ru 3$d$
intensity increases gradually with the increase in Sr content in the
sample as expected. The same spectral region collected with Mg
$K\alpha$ radiations exhibit significantly different intensity ratio
between Sr 3$p$ and Ru 3$d$ signals although the spectral lineshape
seem to be similar at both the photon energies in a particular
composition. This is shown by superimposing the Mg $K\alpha$ spectra
over the Al $K\alpha$ spectra in Fig. 1(a). Clearly the integral
intensity of the Ru 3$d$ signal exhibit enhanced intensity in the Mg
$K\alpha$ spectra relative to the intensity of the Sr 3$p$
intensity. Two reasons can be thought for such intensity change -
(i) photoemission cross section related change in intensity. The
photoemission cross section\cite{yeh} of Ru 3$d$ and Sr 3$p$ states
are very similar at these two photon energies and hence can be ruled
out here. (ii) The other possibility is the change in surface
sensitivity - the surface sensitivity of the technique increases
with the decrease in photon energy.\cite{CSVO-Manju} The enhancement
of Ru 3$d$ signal intensity with the increase in surface sensitivity
suggests that the sample surface is dominated by Ru contributions.
Thus, the oxygen coordination of surface Ru will be different from
the bulk ones leading to different electronic structure, which is
consistent with the earlier observations.\cite{prbr-ravi}

\begin{figure}
 \centerline{\epsfysize=5.0in \epsffile{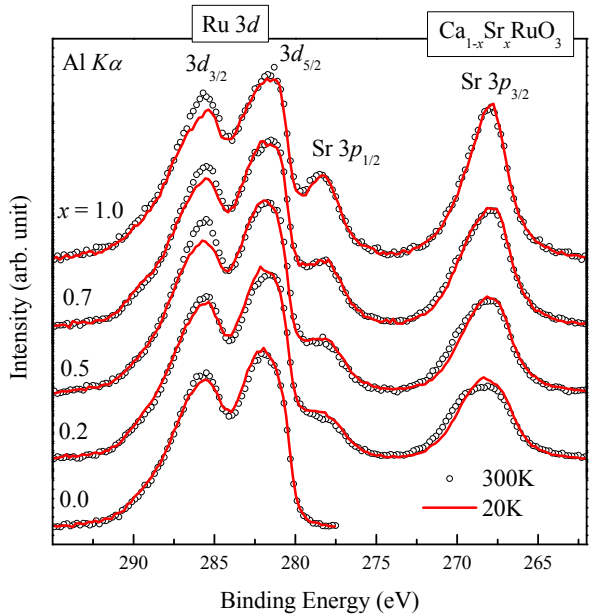}}
\vspace{-24ex}
 \caption{Ru 3$d$ and Sr 3$p$ core level spectra in Ca$_{1-x}$Sr$_x$RuO$_3$ for
different values of $x$ collected using Al $K\alpha$ photon energy
at 300 K (open circles) and 20 K (solid lines).}
\end{figure}

In Fig. 2, we show the Ru 3$d$ and Sr 3$p$ spectral functions
collected at 300 K and 20 K using Al $K\alpha$ radiations.
Interestingly, the Sr 3$p$ lineshape becomes narrower at 20 K
compared to the 300 K spectra keeping the overall intensity of the
features unchanged. However, Ru 3$d$ spectra exhibit decrease in
intensity in the 3$d_{3/2}$ spectral region relative to the
intensity in the 3$d_{5/2}$ region. These spectral evolutions are
unusual as the relative intensity ratio depends on the multiplicity
of the corresponding levels - the multiplicity is not dependent on
temperature. The relative spectral intensity change is possible only
if there is additional contribution presumably due to 3$d_{5/2}$
satellite appearing in the 3$d_{3/2}$ spectral region and the
satellite to main peak intensity ratio changes. Such possibility is
also manifested by a shoulder at about 287.5 eV, which can be
attributed to a satellite corresponding to the 3$d_{3/2}$
photoemission signal. The spectral change appear to be most
significant in the magnetically ordered compounds (SrRuO$_3$ end)
compared to the paramagnetic compositions (CaRuO$_3$ end). The bulk
valence band spectra also showed significant evolution with
temperature in the magnetically ordered compounds, while the
non-ordered compositions remained unchanged with
temperature.\cite{CSRO_APL} Magnetic ordering sets in due to the
inter-site exchange interactions and thus, the electronic states
mediating the inter-site exchange interactions are strongly coupled
to the Ru-moment making it easier to screen the core hole in the
photoemission final state that leads to larger intensity of the well
screened feature.

\begin{figure}
 \centerline{\epsfysize=5.0in \epsffile{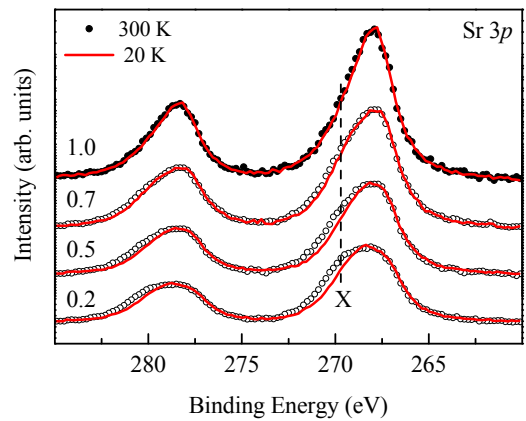}}
\vspace{-42ex}
 \caption{Sr 3$p$ core level spectra of Ca$_{1-x}$Sr$_x$RuO$_3$ for
different values of $x$ collected using Al $K\alpha$ photon energy
at 300 K (open circles) and 20 K (solid lines). The Sr 3$P_{1/2}$
spectral shape is approximated to be similar to the 3$p_{3/2}$
lineshape.}
\end{figure}

Sr 3$p$ spectra are shown in Fig. 3. Normalization by the intensity
of the most intense feature exhibits narrowing of the spectral
lineshape in the intermediate compositions while the end member,
SrRuO$_3$ possess identical and sharp lineshape at both the
temperatures studied. There is an additional feature marked by 'X'
in the figure in the intermediate compositions that makes the
features broad. It has been found that Sr-O covalency is weaker than
Ca-O covalency\cite{CSRO-band} that led to stronger distortion of
the GdFeO$_3$ type orthorhombic distortion in CaRuO$_3$. Thus, the
Madelung potential of Sr surrounded by both Sr and Ca sites will be
enhanced due to larger local distortion (the apical oxygens come
closer to the Sr sites) relative to the Madelung potential at the Sr
sites surrounded by only Sr ions. This explains appearance of an
additional feature in the intermediate compositions at higher
binding energy. Interestingly, decrease in temperature leads to a
narrowing of the peaks - the two features come closer to each other
presumably due to the weaker distortion in the thermally compressed
structure at low temperatures.

\begin{figure}
 \centerline{\epsfysize=5.0in \epsffile{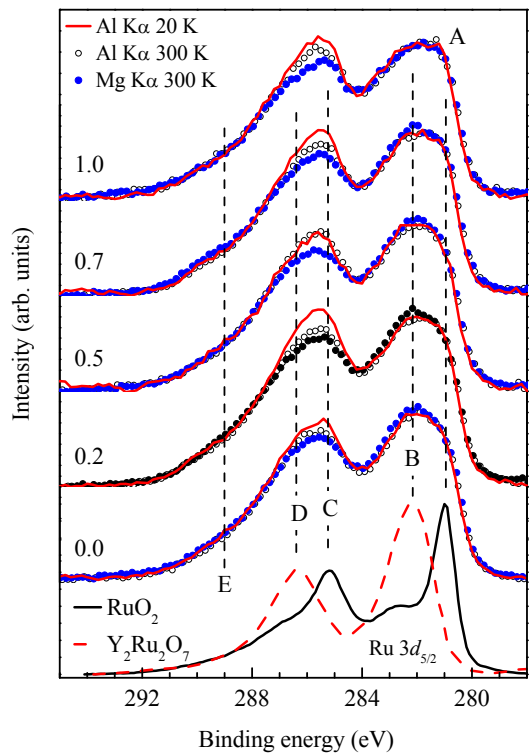}}
\vspace{-12ex}
 \caption{Ru 3$d$ core level spectra of Ca$_{1-x}$Sr$_x$RuO$_3$ for
different values of $x$ collected using Al $K\alpha$ (open circles)
and Mg $K\alpha$ (solid line) photon energies at 300 K and Al
$K\alpha$ photon energy at 20 K (solid symbols). The Sr 3$P_{1/2}$
spectral contributions are subtracted out.}
\end{figure}

Ru 3$d$ spectra extracted by subtracting the Sr 3$p_{1/2}$
contributions at the lower binding energy tail are shown in Fig. 4
for different photon energies and temperatures. All the spectra are
normalized by the intensity of the most intense peak. Signature of
five distinct features denoted by A, B, C, D and E are observed in
all the spectra. While the two energy regimes below and above 284 eV
binding energy can be attributed broadly to 3$d_{5/2}$ and
2$d_{3/2}$ spin-orbit split signals, the two feature structure of
each of the spin-orbit split components have been observed in many
other ruthanates and shown to be associated with different final
state effects in photoemission process arising due to screening of
core hole by valence electrons
\cite{cox-jpc-1983,kim-prl-2004,takizawa-prbr-2005}.

The relative intensity of the feature A and B does not change with
the change in photon energy and temperature. However, the overall
intensity in the 3$d_{3/2}$ spectral region increases with the
decrease in temperature and reduces with the change in photon energy
towards higher surface sensitivity. These observations suggest the
existence of a second satellite feature corresponding to 3$d_{5/2}$
signal similar to the feature E for 3$d_{3/2}$ signal increases at
low temperatures and the surface spectra has weaker contribution.
Since this feature correspond to another poorly screened feature, it
is reasonable to state that the screening of the core level becomes
less efficient at low temperatures and is stronger at the surface.
In addition, the intensity of the feature A relative to the
intensity of the feature B increases with the increase in $x$ from
$x$~=~0.0 to 1.0 in Ca$_{1-x}$Sr$_x$RuO$_3$.

\begin{figure}
 \centerline{\epsfysize=5.0in \epsffile{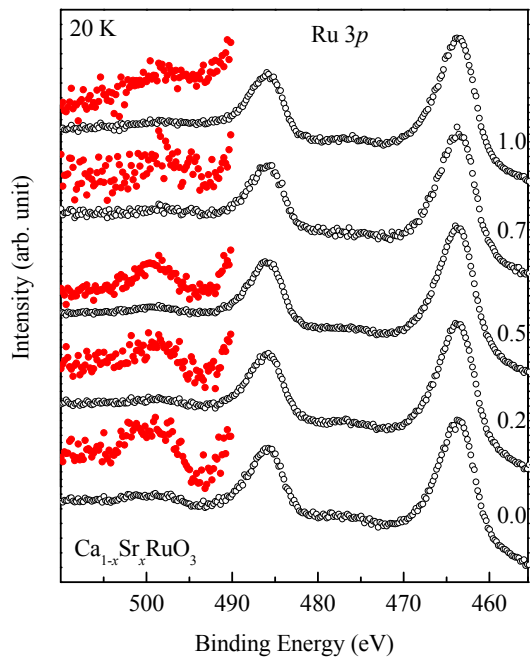}}
\vspace{-18ex}
 \caption{Ru 3$p$  core level spectra (open circles) obtained using Al $K\alpha$ radiations
at 20 K. Satellite corresponding to 3$p_{1/2}$ feature is shown in
an expanded intensity scale (solid circles).}
 \vspace{-2ex}
\end{figure}

In Fig.~5, we show the Ru 3$p$ core level spectra obtained at 20~K
using Al $K\alpha$ radiations. Spectra shows spin orbit split two
peaks at around 463.5~eV and 285.5~eV corresponding to photoemission
signals from Ru~3$p_{3/2}$ and Ru~3$p_{1/2}$ electronic states,
respectively. The spin orbit splitting of about 22~eV and the
lineshape of the peaks are similar for all values of $x$. Weak
satellite features representing the poorly screened final states
appear around 13~eV higher binding energy with respect to the well
screened (main) peak
\cite{bocquet-prb-1992,bocquet-prb-1996,fujimori-prb-1986}. The
relative intensity of the satellite feature with respect to the
intensity of the main peak is very small. Thus, the strength of the
electronic interaction parameters are expected to be significantly
weak in these systems. The relative intensity of the satellite
feature reduces gradually with the increase in $x$; the ratio of the
satellite to main peak intensities is about 20\% in CaRuO$_3$ and
15\% in SrRuO$_3$. A careful look at the satellite signals reveals
that there are two features as shown by rescaling the satellite
feature at around 500~eV. The feature at higher binding energy in
the satellite reduces continuously with the increase in $x$ and
becomes almost invisible in SrRuO$_3$. Understanding of the origin
of such spectral evolutions needs detailed calculations of the core
level spectra. We hope, such calculations including all the
electronic interaction parameters will be carried out in future to
enlighten these interesting effects. Change in temperature does not
have significant influence in the Ru 3$p$ core level spectra.

Electron correlation leads to significant modification of the
valence and conduction bands. The valence band of the correlated
metallic transition metal oxides usually consists of two
features\cite{RMP} - one feature appears at the Fermi level
representing the delocalized electronic states and is known as
\emph{coherent feature}. The other one appears at higher binding
energies derived by the electron correlation strength and represents
the correlation induced electronic states. This feature is termed as
the lower Hubbard band/\emph{incoherent feature}. The electrons
corresponding to the coherent feature are highly mobile and helps to
screen the local positive charge present in the final state of
photoemission. Thus. the core level spectra often consists of at
least two features - one the well screened feature and the other is
the poorly screened feature.

In the Ru 3$d$ spectra, the feature B, corresponds to the poorly
screened core hole final state and the feature A can be attributed
to the final state where the core hole is screened by the conduction
electrons representing the coherent feature. This can be verified by
the Ru 3$d$ core level spectra of Y$_2$Ru$_2$O$_7$ (from Ref.
\cite{cox-jpc-1983}) and RuO$_2$ (from Ref. \cite{kim-prl-2004}) as
shown in Fig. 4. RuO$_2$ is a good metal, where the influence of the
correlation among the valence electron is negligible. On the other
hand, Y$_2$Ru$_2$O$_7$ is a Mott insulator, where all the valence
electrons contribute in the incoherent feature. It is evident from
the figure that the peaks A and C correspond well to the 3$d$ core
level spectrum of RuO$_2$ and the features B and D correspond to the
3$d$ spectra of Y$_2$Ru$_2$O$_7$. The signature of the feature E is
not distinctly evident in the 3$d$ spectra of RuO$_2$ or
Y$_2$Ru$_2$O$_7$ spectra. More studies are required, in particular
the studies based on theoretical calculations to reveal the origin
of this feature.

Interestingly, Ru 3$p$ spectra also exhibit signature of similar
satellite structures in addition to the main peak due to the well
screened final state. The distinct signature of the satellite
features once again reaffirms the scenario based on various
screening channels in these systems. In both, 3$d$ and 3$p$ spectra,
the relative intensity of the satellite peak with respect to the
main peak gradually reduces with the increase in Sr-concentration.
Since, electron correlation strength is quite similar in both the
systems, such spectral evolution can be attributed to the change in
local crystal structure. CaRuO$_3$ possesses orthorhombic distortion
and Ru-O-Ru angle is close to 150 $^o$, and SrRuO$_3$ is less
distorted with Ru-O-Ru angle close to 65$^o$. Increase in Ru-O-Ru
bond angle enhances the hopping interaction parameter and hence
screening of core hole becomes more efficient. The decrease in
satellite intensity with the increase in $x$ in both Ru 3$d$ and
3$p$ spectra can, thus, be attributed to the enhancement of Ru-O-Ru
bond angle in this system.

\section{Conclusions}

In summary, we studied the core level spectra of
Ca$_{1-x}$Sr$_x$RuO$_3$ for various values of $x$ employing
photoemission spectroscopy. The surface appears to be dominated by
Ru-O planes. The Sr core level spectra exhibit multiple feature in
the intermediate compositions, which can be attributed to the
differing Ca-O and Sr-O covalency. The Ru core level spectra exhibit
distinct signature of satellite features due to finite electron
correlation among Ru 4$d$ valence electrons. The intensity of the
satellite feature gradually decreases with the increase in Ru-O-Ru
bond angle. Ferromagnetic compositions shows increase in satellite
intensity at low temperatures that can be attributed to the
depletion of intensity at the Fermi level due to inter-site exchange
coupling leading to long range magnetic order.\cite{CSRO_APL} The
satellite intensity corresponding to the surface Ru core level
spectra are weaker than those in the bulk. All the results manifest
the complexity of the electronic structure due to the interplay
between electron correlation and electron hopping interaction
strength.



\begin{thebibliography}{99}
%
\bibitem{bairo3} G. Cao, J.E. Crow, R.P. Guertin, P.F. Henning, C.C. Homes,
M. Strongin, D.N. Basov, and E. Lochner, Solid State Commun. 113
(2000) 657; K. Maiti, Phys. Rev. B 73 (2006) 115119; K. Maiti, R. S.
Singh, V. R. R. Medicherla, S. Rayaprol and E.V. Sampathkumaran,
Phys. Rev. Lett. 95 (2005) 016404.
%
\bibitem{sr2iro4} B.J. Kim, H. Jin, S.J. Moon, J.-Y. Kim, B.-G. Park,
C.S. Leem, J. Yu, T.W. Noh, C. Kim, S.-J. Oh, J.-H. Park, V.
Durairaj, G. Cao, and E. Rotenberg, Phys. Rev. Lett. 101 (2008)
076402.
%
\bibitem{y2ir2o7} R.S. Singh, V.R.R. Medicherla, Kalobaran Maiti, and E.V.
Sampathkumaran, Phys. Rev. B 77 (2008) 201102(R); K. Maiti, Solid
State Commun. 149 (2009) 1351.
%
\bibitem{sr2ruo4} Y. Maeno, H. Hashimoto, K. Yoshida, S. Nishizaki,
T. Fujita, J. G. Bednorz, and F. Lichtenberg , Nature 372 (1994)
532.
%
\bibitem{nfl} P. Khalifah, I. Ohkubo, H. Christen, and D. Mandrus, Phys. Rev. B
70 (2004) 134426; Y. S. Lee, Jaejun Yu, J. S. Lee, T. W. Noh, T.-H.
Gimm, Han-Yong Choi, and C. B. Eom, Phys. Rev. B 66 (2002)
041104(R).
%
\bibitem{klein} L. Klein, L. Antognazza, T.H. Geballe, M.R. Beasley, and A. Kapitulnik,
Phys. Rev. B, 60 (1999) 1448.
%
\bibitem{rss} R.S. Singh, P.L. Paulose, and K. Maiti,
Solid State Physics: {\it Proceedings of the DAE Solid State Physics
Symposium} 49 (2004) 876.
%
\bibitem{cao} G. Cao, S. McCall, M. Shepard, J.E. Crow, and R.P. Guertin,
Phys. Rev. B, 56 (1997) 321.
%
\bibitem{rana} S. Tripathi, R. Rana, S. Kumar, P. Pandey, R. S. Singh, and D. S.
Rana, Sci. Rep. 4 (2014) 3877; S. Middey, P. Mahadevan, and D. D.
Sarma, Phys. Rev. B 83 (2011) 014416.
%
\bibitem{vidya} R. Vidya, P. Ravindran, A. Kjekshus, H. Fjellvag,
and B.C. Hauback, J. Solid State Chem. 177 (2004) 146.
%
\bibitem{callaghan} A. Callaghan, C.W. Moeller, and R. Ward,
Inorg. Chem. {\bf 5}, 1572 (1966).
%
\bibitem{longo} J.M. Longo, P.M. Raccah, and J.B. Goodenough, J.
Appl. Phys. 39 (1968) 1327.
%
\bibitem{sugiyama} T. Sugiyama and N. Tsuda, J. Phys. Soc. Jpn.
68 (1999) 3980.
%
\bibitem{martinez} J.L. Martinez, C. Prieto, J.
Rodriguez-Carvajal, A. de Andr\'{e}s, M. Valet-Regi, and J.M.
Gonzaler-Calbet, J. Magn. Magn. Mater. 140-144 (1995) 179.
%
\bibitem{prbr-ravi}  K. Maiti and R.S. Singh, Phys. Rev. B 71
(2005) 161102(R).
%
\bibitem{takizawa-prbr-2005} M. Takizawa, D. Toyota, H. Wadati, A. Chikamatsu,
H. Kumigashira, A. Fujimori, M. Oshima, Z. Fang, M. Lippmaa, M.
Kawasaki, and H. Koinuma, Phys. Rev. B 72 (2005) 060404(R).
%
\bibitem{epl-ravi} K. Maiti, R. S. Singh, and V. R. R. Medicherla,
Europhys. Lett. 78 (2007) 17002.
%
\bibitem{prb-debdutta} D. Lahiri, T. Shibata, S. Chattopadhyay, S. Kanungo,
T. Saha-Dasgupta, R.S. Singh, S.M. Sharma, and K. Maiti, Phys. Rev.
B 82 (2010) 094440.
%
\bibitem{spec-core} R.S. Singh and K. Maiti, Phys. Rev. B 76
(2007) 085102; R. S. Singh and K. Maiti, Solid State Comm. 140
(2006) 188.
%
\bibitem{ramarao} M. V. Rama Rao, V. G. Sathe, D. Sornadurai,
B. Panigrahi, and T. Shripathi, J. Phys. Chem. Solids 62 (2001) 797.
%
\bibitem{nakatsugawa} H. Nakatsugawa, E. Iguchi, and Y. Oohara, J.
Phys.: Condens. Mater 14 (2002) 415.
%
\bibitem{kobayashi} H. Kobayashi, M. Nagata, R. Kanno, and Y. Kawamoto,
Mater. Res. Bull. 29 (1994) 1271.
%
\bibitem{yeh} J. J. Yeh, and I. Lindau, At. Data Nucl. Data Tables 32 (1985) 1.
%
\bibitem{CSVO-Manju} K. Maiti, U. Manju, S. Ray, P. Mahadevan, I. H. Inoue,
C. Carbone, and D. D. Sarma, Phys. Rev. B 73 (2006) 052508; K.
Maiti, A. Kumar, D. D. Sarma, E. Weschke, and G. Kaindl, Phys. Rev.
B 70 (2004) 195112.
%
\bibitem{CSRO_APL} R. S. Singh, V. R. R. Medicherla, and K. Maiti, Appl.
Phys. Lett. 91 (2007) 132503.
%
\bibitem{CSRO-band} K. Maiti, Phys. Rev. B 73
(2006) 235110; K. Maiti, Phys. Rev. B 77 (2008) 212407.
%
\bibitem{cox-jpc-1983} P. A. Cox, R. G. Egdellt, J. B. Goodenough, A. Hamnett,
and C. C. Naishj, J. Phys. C: Solid State Phys. 16 (1983) 6221.
%
\bibitem{kim-prl-2004} H. -D. Kim, H. -J. Noh, K. H. Kim, and S. -J. Oh, Phys. Rev.
Lett. 93 (2004) 126404.
%
\bibitem{bocquet-prb-1992} A. E. Bocquet, A. Fujimori, T. Mizokawa, T. Saitoh,
H. Namatame, S. Suga, N. Kimizuka, Y. Takeda and M. Takano, Phys.
Rev. B 45 (1992) 1561.
%
\bibitem{bocquet-prb-1996} A. E. Bocquet, T. Mizokawa, K. Morikawa,
A. Fujimori, S. R. Barman, K. Maiti, D. D. Sarma, and Y. Tokura
Phys. Rev. B 53 (1996) 1161.
%
\bibitem{fujimori-prb-1986} A. Fujimori, M. Saeki, N. Kimizuka,
M. Taniguchi and S. Suga, Phys. Rev. B 34 (1986) 7318.
%
\bibitem{RMP} A. Georges, G. Kotliar, W. Krauth, and M. J. Rozenberg, Rev. Mod.
Phys. 68 (1996) 13; M. Imada, A. Fujimori, and Y. Tokura, Rev. Mod.
Phys. 70 (1998) 1039.
%

\end{thebibliography}
\end{document}